\documentclass[preprint2]{aastex}

\def\sun{\hbox{$\odot$}}
\def\farcm{\hbox{$.\mkern-4mu^\prime$}}
\def\farcs{\hbox{$.\!\!^{\prime\prime}$}}

\slugcomment{Accepted by the Astronomical Journal.}

\shorttitle{Spitzer Imaging of Arp 107}
\shortauthors{Smith et al.}

\begin{document}


\title{Using Spitzer Colors as Diagnostics of Star Formation Regions:
The Interacting Galaxy Arp 107
}


\author{Beverly J. Smith}
\affil{Department of Physics, Astronomy, and Geology, East Tennessee
State University, Johnson City TN  37614}
\email{smithbj@etsu.edu}

\author{Curtis Struck}
\affil{Department of Physics and Astronomy, Iowa State University, Ames IA  50011}
\email{curt@iastate.edu}

\author{Philip N. Appleton}
\affil{Spitzer Science Center, 
California Institute of Technology, Pasadena CA  91125}
\email{apple@ipac.caltech.edu}

\author{Vassilis Charmandaris\footnote{Astronomy Department, Cornell
University, Ithaca NY  14853 and Chercheur Associ\'e, Observatoire
de Paris, F-75014, Paris, France}}
\affil{Department of Physics, University of Crete, Heraklion Greece 71003}
\email{vassilis@physics.uoc.gr}

\author{William Reach}
\affil{Spitzer Science Center, 
California Institute of Technology, Pasadena CA  91125}
\email{reach@ipac.caltech.edu}

\and

\author{Joseph J. Eitter}
\affil{Department of Physics and Astronomy, Iowa State University, Ames IA  50011}
\email{rwl@iastate.edu}



\begin{abstract}
We present Spitzer infrared imaging 
of the peculiar galaxy pair Arp 107, and compare with an optical
H$\alpha$ map and a numerical model of the interaction.
The [3.6] $-$ [4.5] colors of clumps in the galaxy
do not vary around the ring-like primary spiral arm and are consistent with those
of stars, thus these bands are dominated by starlight.
In contrast,
the [5.8 $\mu$m] $-$ [8.0 $\mu$m] colors are consistent
with those of interstellar dust, and
vary by about 0.2 magnitudes around the ring/spiral, with 
redder colors associated with regions with stronger star
formation as indicated by H$\alpha$ and mid-infrared luminosity.
The [4.5 $\mu$m] $-$ [5.8 $\mu$m] colors 
for clumps in this arm
are bluer than dust and redder than stars, 
and vary by 1.3 magnitudes around the arm.
This color is therefore a measure of the relative number of young 
stars to old stars, with a redder color indicating a higher 
proportion of young stars.  
There is a clear azimuthal sequence 
in the [4.5] $-$ [5.8] color around the
arm, indicating a sequence in average stellar age.
The L$_{H\alpha}$/L$_{8.0 {\mu}m}$ ratio varies around the arm
by a factor of $\approx$7; this variation may be due to extinction
or to PAH excitation by non-ionizing photons.
Our model of Arp 107 accounts for the general morphology
of the system, and explains the age variation along the arm
as the result of differences in the time of maximum compression
in the arm.
Using Spitzer colors, 
we are able to distinguish background quasars and foreground 
stars from star forming regions associated with Arp 107.

\end{abstract}





\section{Introduction}

Tidal interactions, collisions, and mergers 
of galaxies play a major role
in the evolution of galaxies, drastically modifying both the morphology
and the star formation rates of galaxies (see review
by \citealp{struck99}).  Interaction-induced star
formation was first investigated by \citet{larson76}
and \citet{struckmarcell78}, who provided evidence
that interacting 
galaxies have a larger scatter in star formation properties
than isolated galaxies.  The Infrared
Astronomical Satellite (IRAS) 
revealed a new population of high
infrared luminosity galaxies with large star formation rates
\citep{soifer87, smith87}
that are the result of mergers 
\citep{sanders88}.
Mid-infrared observations with the Infrared
Space Observatory
(ISO) suggested that the mid-infrared is a good
measure of the global star formation rate in galaxies
\citep{roussel01, forster04}.
The mid-infrared emission is dominated by broad dust emission
features, which are believed to be due to polycyclic aromatic
hydrocarbons (PAHS) \citep{leger84, allamandola85}.
With 
the advent of the Spitzer infrared telescope 
\citep{werner04}, higher angular
resolution mid-infrared imaging of galaxies is now possible,
making feasible the detailed study of individual
star formation complexes in
interacting systems.

In this paper, we present Spitzer infrared images
of the interacting galaxy pair Arp 107 (UGC 5984), and compare
with an optical H$\alpha$ map to investigate
the properties of star forming clumps with the galaxy.
We also present a new numerical model of the interaction.
The larger galaxy in the pair
has a Seyfert 2 nucleus \citep{keel85}, 
a strong primary spiral arm with a
partial ring-like structure, and a tidal tail
(\citealp{arp66}; see Figure 1).
It is connected to the smaller companion by a bridge.
Arp 107 
is at a distance of 138 Mpc \citep{keel85}, asssuming
H$_o$ = 75 km~s$^{-1}$~Mpc$^{-1}$.

Globally, Arp 107 has a relatively low normalized star formation rate
compared to other interacting galaxies.  Its 
far-infrared luminosity (40 $-$ 120 $\mu$m) 
is 1.1 $\times$ 10$^{10}$ L$_{\sun}$,
similar to the median for other Arp Atlas galaxies,
however, 
it has a far-infrared to blue luminosity ratio 
L$_{FIR}$/L$_B$ = 1.6 and an IRAS 60-to-100 $\mu$m flux ratio
F$_{60 {\mu}m}/F_{100 {\mu}m}$ = 0.19,
at the low end of the range for Arp
galaxies \citep{bushouse88}.
Arp 107 is relatively rich in gas, with
a total HI mass of 1.2 $\times$ 10$^{10}$ M$_{\sun}$
\citep{bushouse87}
and a molecular gas mass of 9.6 $\times$ 10$^{9}$ M$_{\sun}$
in a 22$''$ beam
\citep{zhu99},
assuming the standard Galactic I$_{CO}$/N$_{HI}$ ratio
\citep{bloemen86}.

\section{Observations}

\subsection{Spitzer Observations}

Images of Arp 107 were made
on December 17, 2004 in
the 3.6, 4.5, 5.8, and 8.0 $\mu$m
broadband
filters of the
Spitzer
Infrared Array Camera (IRAC;
\citealp{fazio04}).
The full bandwidths of these filters at 50\% of the average in-band
transmission are 0.75 $\mu$m, 1.01 $\mu$m, 1.42 $\mu$m, and 2.93 $\mu$m,
respectively
\citep{fazio04}.
A total of 23 dithered exposures of
12 seconds each were made
per filter, giving
a total integration time
per filter 
of 276 seconds.
The standard post-pipeline 
Basic Calibrated Data (post-BCD)
mosaicked images were used for
this analysis, providing a platescale of 1\farcs2 per pixel.
The FWHM of the point spread function at 3.6, 4.5, 5.8, and 8.0 $\mu$m
are 1\farcs7, 1\farcs7, 1\farcs9, and 2\farcs0, respectively
\citep{fazio04}.
The total field of view mapped with IRAC is 9\farcm0 $\times$ 14\farcm8,
with the long axis at a position angle of 19.8$^{\circ}$ east of north.
For 3.6, 4.5, 5.8, and 8.0 $\mu$m, 
the rms noise levels in the final maps
are 0.0072, 0.0099, 0.041, and 0.041 MJy~sr$^{-1}$,
respectively.

Additional observations of Arp 107
were made 
on 
December 3, 2004, 
using 
the 
24 $\mu$m filter (bandwidth $\approx$ 5 $\mu$m)
of
the 
Spitzer
Multiband Imaging Photometer
(MIPS;
\citealp{rieke04}).
Two 10-second cycles were
made with the standard photometry AOT, giving an effective
exposure of 312 seconds.
The pixel size of the post-BCD images is 2\farcs45,
and the FWHM of the point spread function is 6$''$.
The
total field of view mapped is 7\farcm4 $\times$ 8\farcm1, with
the long axis of the field of view at a position angle of 23.9$^{\circ}$
east of north.
The final rms noise level in this map is 0.048 MJy~sr$^{-1}$.

\subsection{Optical Observations}

Optical images of Arp 107 were obtained with the 0.6 meter Erwin
Fick telescope in Boone Iowa on 2005 April 8, 14, and 15, 
under clear dark skies.
An 1024 $\times$ 1024 Andor DU434-BV CCD was used with
a pixel size of 1\farcs05, giving a field of view of 17\farcm9.
A total of seven 120 second exposures were made in a broadband R
filter,
along with 7 $\times$ 450 second images in a redshifted
H$\alpha$ filter centered at 6814\AA ~with FWHM 73\AA.
For Arp 107, this filter contains
both H$\alpha$ and the [N~II] $\lambda$6583 line.
The data were reduced in the standard way using the Image
Reduction and Analysis Facility 
(IRAF\footnote{IRAF is distributed by the National 
Optical Astronomy Observatories, which are operated 
by the Association of Universities for Research 
in Astronomy, Inc., under cooperative agreement with the National
Science Foundation.})
software.  Continuum subtraction was accomplished using the scaled 
R band image.
The standard star Feige 56 was observed for H$\alpha$ calibration.
The H$\alpha$ calibration is estimated to be accurate to $\approx$30$\%$.

\section{RESULTS}

\subsection{The Optical Images}

The final co-added Fick R band image of Arp 107 is presented in Figure 1.
In this image, the strong primary arm/ring-like structure
is apparent
in the south, 
which connects to a tidal
tail to the northwest.
A bridge connects the main galaxy to a compact
companion to the northeast.  The bridge
appears double in this image, and a
plume is visible to the northeast of the companion.

The continuum-subtracted H$\alpha$+[N~II] map is shown in Figure 
2.
Seven star formation regions were detected in the ring/arm,
in addition to the Seyfert nucleus.
The H$\alpha$+[N~II] luminosities of these HII region complexes
range from $\approx$4 $\times$ 10$^{39}$ erg~s$^{-1}$ to
3 $\times$ 10$^{40}$ erg~s$^{-1}$ in 5$''$ circular apertures
(see Table 1).
These are similar to those of the most luminous HII regions
in ring galaxies (\citealp{marston95}; J. Higdon 2004, private
communication).
The total H$\alpha$+[N~II] luminosity for this system, excluding
the Seyfert nucleus, is $\approx$1.1 $\times$ 10$^{41}$ erg~s$^{-1}$,
at the lower end of the range for the ring galaxies in the \citet{marston95}
sample.  For the Seyfert nucleus, 
L$_{H{\alpha}+[N~II]}$ $\approx$ 2 $\times$
10$^{41}$ erg~s$^{-1}$, consistent with the \citet{keel85} measurements.
The 3$\sigma$ 
L$_{H{\alpha}+[N~II]}$ upper limit for the companion is $\approx$ 4 $\times$
10$^{40}$ erg~s$^{-1}$.
Approximately correcting for [N~II] in the filter and excluding
the Seyfert nucleus, the far-infrared to H$\alpha$ luminosity ratio
for Arp 107 is L$_{FIR}$/L$_{H{\alpha}}$ $\approx$ 520.  This is
typical of interacting galaxies \citep{bushouse87}.

\subsection{The Spitzer Images}

The Spitzer images
at the five wavelengths
are presented in Figure 3.
The morphology of this system varies with
wavelength.
The bridge is clear at 3.6 $\mu$m but almost disappears by 8.0 $\mu$m.
In contrast,
the knots in the northwestern part of the ring/arm are more prominent
at 8.0 $\mu$m than at 3.6 $\mu$m.
The companion is very faint at 24 $\mu$m but the
ring/arm is clearly seen.
The double bridge structure is visible in the two shorter wavelength
IRAC bands.

At 3.6, 4.5, 5.8, 8.0, and 24 $\mu$m, the total flux densities for the larger
galaxy in the pair are 21.7 mJy, 14.2  mJy,
19.9 mJy, 38.0 mJy, and 53.4 mJy, 
respectively.
For the smaller galaxy, the total flux densities are 13.2
mJy, 8.5 mJy, 13.0 mJy, 4.4 mJy,
and 0.48 mJy,
respectively.  
Absolute calibration uncertainties are $\le$10$\%$ (IRAC Data Manual;
MIPS Calibration Manual).  Statistical uncertainties are $\le$1$\%$,
except for the companion at 8.0 and 24 $\mu$m (2$\%$ and 17$\%$, respectively).
Ambiguities in selecting the boundaries of the galaxies and sky subtraction
introduce an additional uncertainty of $\le$20$\%$ in most
cases, 40$\%$
at 5.8 $\mu$m, and 60$\%$ for the companion at 24 $\mu$m.

The global 
L$_{H\alpha}$/L$_{8.0 {\mu}m}$ and 
L$_{H\alpha}$/L$_{3.6 {\mu}m}$ 
ratios are $\approx$0.0083 and 0.012, 
respectively,
for the main galaxy (excluding the Seyfert nucleus), and $\le$0.025
and $\le$0.0061 for the companion.
For comparison,
the 
global 
log L$_{H\alpha}$/L$_{8.0 {\mu}m}$ 
and
L$_{H\alpha}$/L$_{3.6 {\mu}m}$ 
ratios for the
spiral galaxy NGC 7331 
are $\approx$0.013 and 0.030, respectively,
calculated using the \citet{young96} H$\alpha$ measurement
and the Spitzer flux densities of \citet{regan04}.

\subsection{IRAC Colors of Star Forming Clumps}

In the 8 $\mu$m image, 29 
clumps were identified by eye in the vicinity
of Arp 107.  These are labeled 
on the 8 $\mu$m image in Figure 4, 
encircled by 5$''$ radii regions, and are listed in Table 1.
Note that some fainter clumps are not included.
When a second fainter clump was very close to
a brighter clump so that their regions strongly overlapped,
the second clump was excluded.

Photometry of these clumps was done using the IRAF 
{\it daophot} routine,
with 4.1 pixel (5$''$) radii apertures (Table 1).
For background subtraction, the local galaxian background 
was determined using the
mode in an annulus surrounding the source, with an inner radius
of 5 pixels (6$''$) and an outer radius of 10 pixels (12$''$).
To estimate the uncertainty in the Spitzer colors of the clumps
due to background subtraction,
we experimented with varying the method of determining the sky.
In addition to the statistical uncertainties determined
from the rms in the background annuli, in calculating the colors
we added in quadrature a
second uncertainty term,
determined from comparing the 
clump colors obtained with the above method with those
obtained with slightly larger inner and outer
annulus radii of 7 pixels (8\farcs4)
and 12 pixels (14\farcs4).
This additional term increases the median size of the final
uncertainties in the [3.6] $-$ [4.5], [4.5] $-$ [5.8],
[5.8] $-$ [8.0], and [3.6] $-$ [8.0] colors
by 15$\%$, 33$\%$, 8$\%$, and 66$\%$, respectively.

No color corrections 
were made in determining the IRAC colors.
Aperture corrections were included, although they 
have only a tiny effect on the IRAC colors
of these clumps.
As given in the IRAC Data Manual,
the aperture corrections 
are 1.06, 1.06, 1.07, and 1.09 for IRAC bands 1 $-$ 4, thus 
the changes to the IRAC colors due to
the aperture corrections are negligible.
These clumps are unresolved or barely resolved at
the resolution of the mosaicked images 
(2\farcs4 $-$ 3\farcs6,
compared with 2\farcs0 $-$ 2\farcs4
for bright point sources in the field).

In Figures 5 and 6, the IRAC colors of the brightest 8 $\mu$m
clumps in Arp 107 are plotted (black crosses) and identified.
Both statistical and background selection uncertainties
are included in these plots, but absolute calibration
uncertainties ($\le$10$\%$; IRAC Data Manual) are not included.
Figures 5 and 6 also contain
the colors of
M0III stars (open dark blue square)
(from M. Cohen 2005, private communication)
and the mean colors for the field stars in \citet{whitney04}
(magenta triangle).   The colors of normal stars all lie
within $\approx$0.5 
magnitudes of 0,0 on these plots (M. Cohen 2005, private
communication).
The green diamonds in these figures mark the predicted
IRAC colors for interstellar dust from \citet{li01} (see Table 5 in
that paper), for 
a very large range of interstellar radiation field (ISRF) intensities
(from 0.3 $-$ 10,000 $\times$ that in the solar neighborhood).
Note that the IRAC colors of the \citet{li01}
dust model varies very little with ISRF.
The red circles in Figure 5
show the locations of the Sloan Digitized Sky Survey
quasars in the Spitzer Wide-Area Infrared Extragalactic Survey (SWIRE)
Elais N1 field \citep{hatz05}.
These quasars have
redshifts between 0.5 and 3.65; since their spectral energy distributions
are power laws, their infrared colors do not vary much with redshift.

The colors of Source \#23 (the bright source to the west of the
galaxy) are similar to those of the quasars.  Thus this is likely a background
quasar, unassociated with Arp 107.
The nucleus of the companion, source \#28, is very
blue, with stellar colors 
at all wavelengths.  Thus it contains
almost all old stars and little
ISM. 
This is consistent with its absorption-line optical spectrum (W. Keel 2005, 
private communication).
The nucleus of the primary galaxy, \#14, is also quite blue,
suggesting an old stellar population, but not as old as the companion
nucleus \#28.
This nucleus also has some interstellar matter, as shown by its large
[5.8] $-$ [8.0] color.
Source \#9,
a concentration inside the radius of the ring, also has 
colors indicative of an old star, suggesting it is a foreground
star.
This is confirmed by its optical spectrum, which is that of
an M1$-$2V star (W. Keel 2005, private communication).

The clumps in the upper part
of Figure 6 have
[5.8] $-$ [8.0] colors close to
that predicted by the \citet{li01}
dust model, indicating
that
these wavelengths are dominated
by dust.  
The 8.0 $\mu$m IRAC band
includes the strong 7.7 and 8.6 $\mu$m
PAH emission features as well as a dust
continuum from transiently-heated very small grains, while the 
5.8 $\mu$m IRAC band contains the 5.7 $\mu$m PAH
feature \citep{li01}.
In contrast,
the [4.5] $-$ [5.8] 
colors of the Arp 107 clumps
are much bluer than those of dust, 
thus
there are 
stellar contributions at 4.5 $\mu$m.
The [3.6] $-$ [4.5] colors of these clumps
are similar to those of stars (Figure 5), indicating
that these bands are dominated by starlight.

In Figures 7 and 8, we have plotted the IRAC colors of the 
fainter
Arp 107 clumps.
The colors of
Source \#19, 
off the western side
of the galaxy, indicate that it is likely
a background quasar.
Sources \#1, \#2, and \#3 
(south of the ring), \#12 (west of the ring),
\#27 (in
the bridge), \#29
(at the
northern tip of the tail),
and \#25 (northwest
of the ring)
have [3.6] $-$ [4.5] and
[4.5] $-$ [5.8] colors like old stars, but 
have
excess emission in the 8.0 $\mu$m
band compared to
old stars. 
This suggests that
either these are very old star formation regions associated with the
galaxy, with some PAH emission but mainly cooler stars, or
they are foreground stars, and the 8 $\mu$m band is contaminated
by extended dust emission from Arp 107.

\subsection{An Azimuthal Color Gradient Around the Ring/Arm}

In the
[4.5] $-$ [5.8] vs. [5.8] $-$ [8.0] plot for the bright clumps (Figure 6),
there is a sequence of sources
at an almost
constant [5.8] $-$ [8.0] color of $\approx$2, with
increasing [4.5] $-$ [5.8] 
colors.   The sequence from left to right at the top of 
Figure 6
is a sequence in increasing ratio of young stars/old stars.
This sequence,
\#4, \#5, \#7, \#10, \#21, and \#26,
is the order these clumps
appear in the ring/arm, going in a
counterclockwise direction
from \#4 to \#26.
The northern part of the ring/arm (clumps \#21 and \#26) has
a younger stellar population than the 
southern part (\#4 and \#5) on average.
This trend in the [4.5] $-$ [5.8] color
is better shown in Figure 9, plots of position angle (east of north)
vs.\ the IRAC colors. 
There is a strong azimuthal variation in the [4.5] $-$ [5.8] color
of $\approx$1.3 magnitudes.
In contrast, no variation is visible
in the [3.6] $-$ [4.5] color around the ring/arm.
The approximately constant [3.6] $-$ [4.5] color of $\approx$ 0.0 around
the ring/arm suggests that these bands are dominated by emission by cool stars.

In the 
[5.8] $-$ [8.0] color, the variation is small (0.2 magnitudes) compared
to that in [4.5] $-$ [5.8], and the northern clumps are redder
in [5.8] $-$ [8.0] than the southern clumps.
This small variation 
and the similarity to the colors of interstellar dust indicates
that these two bands are dominated by dust emission, and the dust
spectrum does not vary strongly around the ring/arm.
The [5.8] $-$ [8.0] color increases in the north, in the regions
where the H$\alpha$ and mid-infrared luminosities get larger (see
Section 3.6).  Thus this color gets redder as the ISRF gets
stronger, as predicted by theoretical models
of interstellar dust (see Table 5 in \citealp{li01}).  
As the ISRF increases, the underlying mid-infrared continuum
from very small grains increases, contributing to the 8 $\mu$m IRAC
band, and producing redder [5.8] $-$ [8.0] colors.

Figure 9 shows a large variation in the [3.6] $-$ [8.0] color
around the ring/arm (bottom panel), which is mainly due to changes
in the [4.5] $-$ [5.8] color (second panel).

\subsection{The IRAC-MIPS Colors of Clumps}

From the MIPS 24 $\mu$m map, 24 $\mu$m fluxes were measured using
a 5$''$ radius aperture.  The background was determined using
the mode in 
an annulus with an inner radius of 10$''$ and an outer radius of 15$''$. 
The 5$''$ radius aperture is too small to include
all of the light from the clump, given the $\approx$6$''$ FWHM PSF for MIPS
24 $\mu$m observations and the 2.45$''$/pixel plate scale.
Using a larger aperture radius, however, is not a viable option, since
the clumps are crowded.  
Thus a substantial aperture correction is required.
According to the MIPS Data Handbook, a 5$''$ aperture includes only
$\approx$50$\%$ of the total light from a point source.
Testing this using large aperture data for the isolated source
\#23 (which may be a background quasar; see Section 3.3) and two bright 
sources in the field far from the galaxy, we get consistent
aperture corrections.  
This correction and the 8 $\mu$m 
aperture correction given above increases
the [8.0] $-$ [24] colors by
0.65 magnitudes.
No color corrections were included in determining these colors.
To estimate the uncertainty in the colors due to the selection of background
regions, we also calculated colors with a smaller background annulus
with inner radius of 5$''$ and outer radius of 10$''$, and compared
with the original colors.  This additional color uncertainty was added
in quadrature to the statistical uncertainty.  This extra factor
increases the
sizes of the 
errorbars on the IRAC-MIPS colors of the clumps by a factor of 
$\approx$2.7
over the statistical errors.

In Figure 10,
the [5.8] $-$ [8.0] vs. [8.0] $-$ [24] aperture-corrected
colors for the clumps are plotted.
The green diamonds show the locations of
the \citet{li01} dust models for various ISRFs.
The red circles are the locations of the 
\citet{hatz05} quasars.
Stellar photospheres are expected to lie near
[5.8] $-$ [8.0] $\approx$ 0, [8.0] $-$ [24] $\approx$ 0.

The foreground star \#9 
again has colors similar to those of stars, as expected.
The [8.0 $\mu$m] $-$ [24 $\mu$m] 
color of Source \#23, like its IRAC colors, is similar to those of the quasars,
further supporting the hypothesis that it is a background quasar.
Sources \#1 and \#12 are also in the vicinity of the quasars.
In this plot, the Seyfert nucleus, clump \#14, has colors similar
to those of quasars, with a very red [8.0] $-$ [24] color.
The nucleus of the companion, \#28, shows a small excess relative to the 
expected [8.0] $-$ [24] color of stars ($\approx$0.0), 
thus there may be a small amount of 
cold dust associated with this nucleus.

The [8.0] $-$ [24] colors
of the brightest ring/arm clumps, \#10, \#21, and \#26,  
are similar to those expected for
standard dust models with
weak radiation fields (approximately solar neighborhood).
This is consistent with their relatively low observed average
5 $-$ 8.5 $\mu$m surface brightnesses of $\approx$10
L$_{\sun}$~pc$^{-2}$, which corresponds to an ISRF of $\approx$2
times that in the solar neighborhood, according to \citet{forster04}.
In Figure 11,
the [8.0] $-$ [24] colors for the ring/arm clumps are plotted
as a function
of position angle. 
A possible trend is seen, in that the clumps near the southern part
of the ring/arm,
\#4 $-$ \#7, 
have slightly bluer 
[8.0] $-$ [24] 
colors than those in the northwest,
however, this is uncertain.
If this is confirmed, it would imply that the ISRF is stronger
in the northern part of the ring/arm, as also indicated
by the IRAC colors.

\subsection{H$\alpha$ to Mid-Infrared Ratios}

Of the extra-nuclear star formation regions in Arp 107,
only clumps \#4, \#5, \#7, \#10, \#16, \#21, and \#26 were 
detected in H$\alpha$, with clumps \#4 and \#5 detected at the 5$\sigma$
level
(Section 3.1; Table 1).  In the top panel of Figure 12, for the selected
clumps we compare the ratio of the H$\alpha$
luminosity L$_{H\alpha}$ to the 3.6 $\mu$m luminosity L$_{3.6 {\mu}m}$ with the
8.0 $\mu$m magnitude, correcting for [N~II] in
the H$\alpha$ filter assuming [N~II]/H$\alpha$ $\approx$ 0.3.  
In the second panel of Figure 12, we plot
L$_{H\alpha}$/L$_{8.0 {\mu}m}$ vs. the 8.0 $\mu$m magnitude.
For the clumps detected in H$\alpha$, 
log L$_{H\alpha}$/L$_{8.0 {\mu}m}$ $\approx$ $-$1.3 to $-$2.2.

Figure 12 shows that 
clumps \#4, \#5, \#6, and \#7 are deficient in 
observed H$\alpha$ compared to clumps \#10, \#21, and \#26.
The ratios
L$_{H\alpha}$/L$_{8.0 {\mu}m}$ 
and
L$_{H\alpha}$/L$_{3.6 {\mu}m}$ vary systematically around the ring/arm
(Figure 13), with 
L$_{H\alpha}$/L$_{8.0 {\mu}m}$ $\approx$ 2 $-$ 11 times larger
in the north and west compared to the south and east.

One possible explanation for the low 
L$_{H\alpha}$/L$_{8.0 {\mu}m}$ ratios in clumps \#4 $-$ 7 is
that these clumps have some contributions to
their 8 $\mu$m emission from starlight.
However, the fact that their [5.8] $-$ [8.0] colors are
similar to those of interstellar matter and also to the other 
clumps in the ring/arm indicates that this is not the case.

A second hypothesis is that in clumps \#10, \#21, and \#26, 
strong radiation
fields may be destroying the PAH molecules contributing to
the 8.0 $\mu$m Spitzer data, causing an increase
in L$_{H\alpha}$/L$_{8.0 {\mu}m}$.
Using ISO data for a large sample of spiral and starburst 
galaxies, \citet{forster04} found
a good correlation between the 
mid-infrared
luminosity and inferred star formation rates for 
5 $-$ 8.5 $\mu$m surface brightnesses below $\approx$10$^4$ 
L$_{\sun}$~pc$^{-2}$;
above this surface brightness,
the mid-infrared appeared weaker than expected,
suggesting PAH destruction.
In Arp 107, however,
such large-scale PAH destruction is not expected to happen, since
the radiation fields are not intense enough.
For the brightest clumps in the Arp 107 ring/arm, the
5 $-$ 8.5 $\mu$m surface brightnesses of $\approx$10 
L$_{\sun}$~pc$^{-2}$ are well below
this threshold, thus large-scale PAH destruction is not expected.
Furthermore, the [5.8] $-$ [8.0] colors
of clumps \#21 and \#26 are 
slightly
redder than those of \#4 and \#5, opposite to what would be expected if
the 8 $\mu$m broadband flux is being depressed due to destruction of PAHs by
an intense UV field.

A third possibility is that non-ionizing photons in the southern part
of the ring/arm contribute significantly to heating the PAH 
molecules,
and therefore the 8 $\mu$m band is not a perfect tracer of star formation.
PAH excitation by non-ionizing photons
has been noted in reflection nebulae \citep{sellgren90, uchida98,
uchida00, li02}, and has been suggested as a contributor to the 
global mid-infrared of spiral galaxies 
based on ISO data
\citep{boselli04}.
The PAH features may therefore be a tracer of B stars as well
as O stars \citep{peeters04}.
If the southern and eastern part of the ring/arm contains a significant
post-starburst population, these moderately-young stars may be
sufficient to excite
the PAH molecules and produce the observed mid-infrared emission.  
The bluer [4.5] $-$ [5.8], [5.8] $-$ [8.0], and [8.0] $-$ [24] colors
of the southern clumps \#4 and \#5 compared to clumps \#21 and \#26 
suggest that the mid-infrared continuum from very small grains is weaker
in the south, and therefore the exciting UV field is weaker.  This
is consistent with the hypothesis of an older stellar population
in the south.
This possibility should be investigated further with optical
and infrared 
spectroscopy and stellar
population synthesis modeling.

A fourth
possibility is that the H$\alpha$ emission
is more extincted 
in the clumps in the southern and eastern part of the ring/arm compared to the
northwest.
A relative extinction of A$_V$ $\approx$ 1.5$^{+1.6}_{-0.9}$ magnitudes
around
the ring/arm could
account for the observed deficiency in the H$\alpha$ luminosity.
This possibility should be tested further with 
optical spectroscopic measurements of the Balmer decrement
and/or high angular resolution CO and HI maps.
This will 
determine whether the observed azimuthal trend 
in L$_{H\alpha}$/L$_{8.0 {\mu}m}$
in the ring/arm
is due to extinction or to PAH excitation by non-ionizing photons.

In the L$_{H\alpha}$/L$_{3.6 {\mu}m}$ vs. [8.0] plot,
the same trends are seen: 
the 
L$_{H\alpha}$/L$_{3.6}$ 
ratios for 
clumps \#4, \#5, \#6, and \#7
are lower 
than clumps \#10, \#21, and \#26.
The foreground star \#9 has a very low
L$_{H\alpha}$/L$_{3.6 {\mu}m}$ ratio, as expected.

\section{A Numerical Model of the Encounter}

The morphology of the primary galaxy in Arp 107 consists of a small central
bar-like bulge and a single strong spiral arm with a ring-like shape.
Most of this arm is quite circular, though the
circle is not centered on the bulge and nucleus. Given this, and the fact that
there is material between the companion and the primary, like parts of a bridge,
the system also looks like an asymmetric colliding ring galaxy 
\citep{lynds76, theys77, appleton96}.
On the other
hand, off-center ring waves usually look like ovals, not one-armed spirals
\citep{toomre78, struck90, smith92}.
In addition, our Spitzer maps reveal a strong azimuthal gradient in the ratio of
young stars to old stars around the ring of Arp 107. This asymmetry is also
consistent with the hypothesis that this structure was produced by an off-center
collision between two galaxies. Theoretical models suggest that such encounters
lead to strong azimuthal variations in the gas density and star formation rate
in collisional rings 
\citep{asm, sma, charmandaris93}.

No complete velocity field of Arp 107 is currently available. However, 
\citet{keel96}
obtained a long-slit optical spectrum of Arp 107 along a position angle
of 41$^{\circ}$ east of north. He found that the southwestern side of the main
disk is redshifted compared to the nucleus. The companion has a velocity cz
= 10,572 $\pm$ 
50 km s$^{-1}$, redshifted by 200 km s$^{-1}$ from the main galaxy
(W. Keel 2005, private communication).

To resolve the nature of this collisional system and to better understand the
structure of the star formation in the arm we have attempted to numerically
model it. Given that we have limited kinematical information and
very little information on the gas distribution, we have not attempted to model
it in great detail, but rather to produce a reasonable preliminary model.

\subsection{Details of the Model}

We have used the SPH code described in \citet{struck97}.
The version used here
is essentially the same as that used in \citet{struck03}.
Briefly, this
code uses rigid dark halo potentials for the two galaxies, and hydrodynamic
forces are computed on a grid with fixed spacing. Local gravitational forces are
computed between particles in adjacent cells. 
The galaxy disks are made up of gas
particles and collisionless star particles of equal mass. In these models the
particle numbers were as follows: 34,550 primary disk gas particles, 18,090
companion disk gas particles, 7890 primary star particles and 2490 companion
star particles.

Because the companion appears to be an early type galaxy, its star particles
were initialized with a large range of distances from the central plane and a
large velocity dispersion perpendicular to that plane to give a very thick disk
or bulge-like structure. A substantial gas disk was also added to the companion,
even though there is no evidence for the existence of a gas disk. The goal was
to see how it would be affected by the collision, and whether we would expect it
to be detectable in current or future observations.

We adopt the following scaling constants for the model: time unit = 400 Myr,
length unit = 2.0 kpc, and mass unit = $1.3 \times 10^{10}\ M_{\odot}$. With
these scalings we obtain a peak rotation velocity in the primary of about 170
km/s, which is reasonable within the very limited constraints.

The mass ratio of the two galaxies, determined by comparing the mass of each
within a radius of 10 kpc, is 0.16. The form of the rigid potential used 
gives a
test particle acceleration of,

\begin{equation}
a =
\frac{G{M_h}}{{\epsilon}^2}\
\frac{r/{\epsilon}}{(1 + r^2/{\epsilon}^2)^{n_h}},
\end{equation}

\noindent
where $M_h$ is a halo mass scale, $\epsilon$ is a core radius (set to 4.0 and
2.0 kpc for the primary and companion, respectively), and the index $n_h$
specifies the compactness of the halo. For the primary we use $n_h = 1$, which
gives a flat rotation curve at large radii. For the companion we take $n_h =
1.35$, which gives a declining rotation curve at large radii. This latter choice
seems to produce somewhat better results than two flat rotation curves, but we
have not undertaken a detailed examination of the effects of different values
for these exponents. Because the two galaxies have different halo potentials,
the effective mass ratio (i.e., the ratio of masses contained within radii equal
to the galaxy separation) is large for small separations, and small for large
large separations. This effect makes encounters between the two galaxies quite
impulsive.
The model presented here does not include the effects of dynamical friction, but
other runs included a Chandrasekhar-like frictional term (see \citealp{struck03}).
This term was found to be negligible except for brief times near closest
approach. It had a small effect at the first closest approach, but ultimately
draws the two galaxies into a merger at times past the present (the second close
encounter).

The initial disk sizes were as follows: primary gas disk = 15 kpc, primary star
disk = 12 kpc, companion gas disk = 9.0 kpc, and companion star disk = 3.6 kpc.
These are relatively arbitrary values until 21 cm mapping of the HI
distribution becomes available. The companion disk was initialized in the
computational x-y plane. The primary disk was initialized in the x-y and then
tilted $20^{\circ}$ around the x-axis such that positive y values had positive z
values. After the run the viewing angle was adjusted by rotating the system
$-50^{\circ}$ around the x-axis, $30^{\circ}$ around the y-axis, and
$90^{\circ}$ around the z-axis. The initial position and velocity vectors of the
companion relative to the primary center were: (7.0 kpc, 7.0 kpc, 8.0 kpc) and
(70 km s$^{-1}$, $-$44 km s$^{-1}$, $-$24 km s$^{-1}$).

\subsection{Model Results}

The dashed lines in the top panels of Figure 14 show that, although the companion
collides with the outermost part of the primary gas disk, it then flies a
considerable distance radially out from the disk center and below it (assuming
it began `above' it, i.e., at positive z coordinate values). The model seems to
fit the observed morphology best at a time just before or after the companion
again passes through the primary disk. This is in accord with the significant
line-of-sight velocity difference between the two galaxies.

The lower panels in the figure show that at earlier times there was a
substantial bridge between the two galaxies, and the primary had much more of
the two-armed appearance characteristic of prograde encounters. However, by the
present time, most of the bridge material has fallen onto one or the other
galaxies. What little is left is projected along the line-of-sight, and not very
visible. Some bridge gas originating from the companion continues to fall onto
the primary. This mass transfer was greater in the past, and could plausibly
play a role in feeding the Seyfert nucleus of the primary.

We have run a number of additional exploratory models, and in many of them the
companion has a substantial tidal tail. Depending on initial and orbital
parameters these tails can have many different orientations. In some cases tail
material might account for the debris to the east of the companion and north of
the primary. In the present model, however, debris to the east and west of the
companion consists mostly of stars pulled off the primary disk (see Figure 14).

In these models the primary arm originates as a tidal counter-tail drawn from
primary disk material on the opposite side from
the companion's closest approach point.
The bridge arm originates on the companion side near the beginning of the
models, and its outer part becomes the bridge as described above. The inner part
of the bridge arm is overrun by the primary arm at about the present time in
many of the models. The models suggest that the peculiar `nob' on the eastern side
of the primary arm is a remnant of the bridge arm. In sum, the models suggest
that the one-armed appearance of the primary galaxy is the result of the
asymmetric evolution of an essentially two-armed disturbance. This
asymmetric evolution is the result of substantial m=0 (ring-like) and m=1
(perturbed center) components of that disturbance.

The new Spitzer observations resolve many star-forming clumps in the primary
arm. 
Interestingly, there seem to be
clumps along almost the whole arm, and the star formation is not
significantly stronger in the inner arm regions where one might expect the gas
density to be greater. Moreover, as described above, there is evidence that
clump populations are younger as one tracks the arm outward and around.

The model results in the lower panels of Figure 
14 offer some possible explanations
for these trends. These show that the arm has narrowed and stretched in the last
400 Myr. The overlaps of test particle rings show that compressions increased
greatly on the leading edge of the arm in the west in the first half of this
time period. The figure shows that compressions in the outermost parts of the
arm develop later. The feedback algorithm in the models shows that star-forming
clumps appear rather stochastically at late times in the outer tail, which is
consistent with the test particle results convolved with a random element in the
way gas densities push over a threshold. We suggest that this may be how clumps
\#16, 21 and 29 formed. Alternatively, clumps \#21, 22, 24 and 26 might be the
result of companion debris interacting with the primary disk.

We should emphasize that this relatively straightforward picture of the nature
of the star formation in the primary arm is strongly based on the Spitzer 
infrared colors. The situation would be 
considerably murkier if we had
only H$\alpha$ data to compare to the models.

\section{Conclusions}

We see a strong azimuthal variation
of 1.3 magnitudes
in the [4.5] $-$ [5.8] colors of clumps in the ring/primary arm of
Arp 107.  These colors
are in between those of stars and dust, implying an
azimuthal sequence in the 
ratio of the 
number of old stars to the
number of young stars.
In contrast, no variation is seen in the [3.6] $-$ [4.5] color,
and these colors are consistent with old stars.  Thus these two
shorter wavelength 
bands are dominated by an older stellar population.
Only a small variation of 0.2 magnitudes is seen in the [5.8] $-$ [8.0] color, 
and these colors are consistent with those of interstellar dust,
thus these bands are dominated by interstellar matter.
The [5.8] $-$ [8.0] color becomes redder in regions with
stronger star formation, as indicated by the H$\alpha$ and 
mid-infrared luminosities.

Our numerical model suggests that Arp 107 is the result of a hybrid
collision: part small impact parameter/large angle of attack, as in colliding
ring galaxies, and part prograde planar encounter, as in M51-type bridge/tail
systems. Specifically, the model shows that the strongest compressions occur at
different times along the primary arm. Assuming these compressions trigger star
formation, this differential delay can account for the observed variation of
clump age with azimuth. The Spitzer observations have allowed us to detect this
trend, and the model comparison allows us to understand the driving mechanism.
This example suggests bright prospects for learning about star formation
mechanisms by comparing Spitzer data to models of interacting galaxies.





\acknowledgments

We thank the Spitzer team for making this research possible.
We also thank Bill Keel for providing us with the 
velocity of the companion and the spectral type of the foreground star,
and Martin Cohen for providing stellar colors.
We appreciate helpful suggestions on this project from 
Mark Giroux and the referee Suzanne Madden.
This research was supported by NASA Spitzer grant 1263924 and NSF
grant AST-0097616.
This research has made use of the NASA/IPAC Extragalactic Database (NED) which is operated by the Jet Propulsion Laboratory, California Institute of Technology, under contract with the National Aeronautics and Space Administration.

\clearpage



 \begin{figure}
\caption{
  \small 
The Fick R band image of Arp 107.
Note the strong primary arm/partial ring 
in the south, the tail to the northwest,
the bridge, and the plume to the northeast of the companion.
}
\end{figure}

 \begin{figure}
\caption{
  \small 
The continuum-subtracted H$\alpha$+[N~II] image of Arp 107.
Several star forming regions in the ring/arm are detected, along
with the Seyfert nucleus.
Some artifacts are present near the 
companion due to imperfect continuum subtraction.
}
\end{figure}

 \begin{figure}
\epsscale{.8}
\epsscale{1}
\caption{
  \small 
Left to right, top to bottom:
the Spitzer 3.6, 4.5, 5.8, 8.0, and 24 $\mu$m maps of Arp 107.
The field of view is the same as in Figures 1 and 2.
}
\end{figure}

 \begin{figure}
\caption{
  \small 
The Spitzer 8.0 $\mu$m
map of Arp 107, with 5$''$ radius circles identifying the clumps discussed
in the text.  The clumps are numbered in order by increasing declination (see
Table 1). The number identifying each clump is directly above the 
corresponding circle in this figure.
}
\end{figure}

 \begin{figure}
\plotone{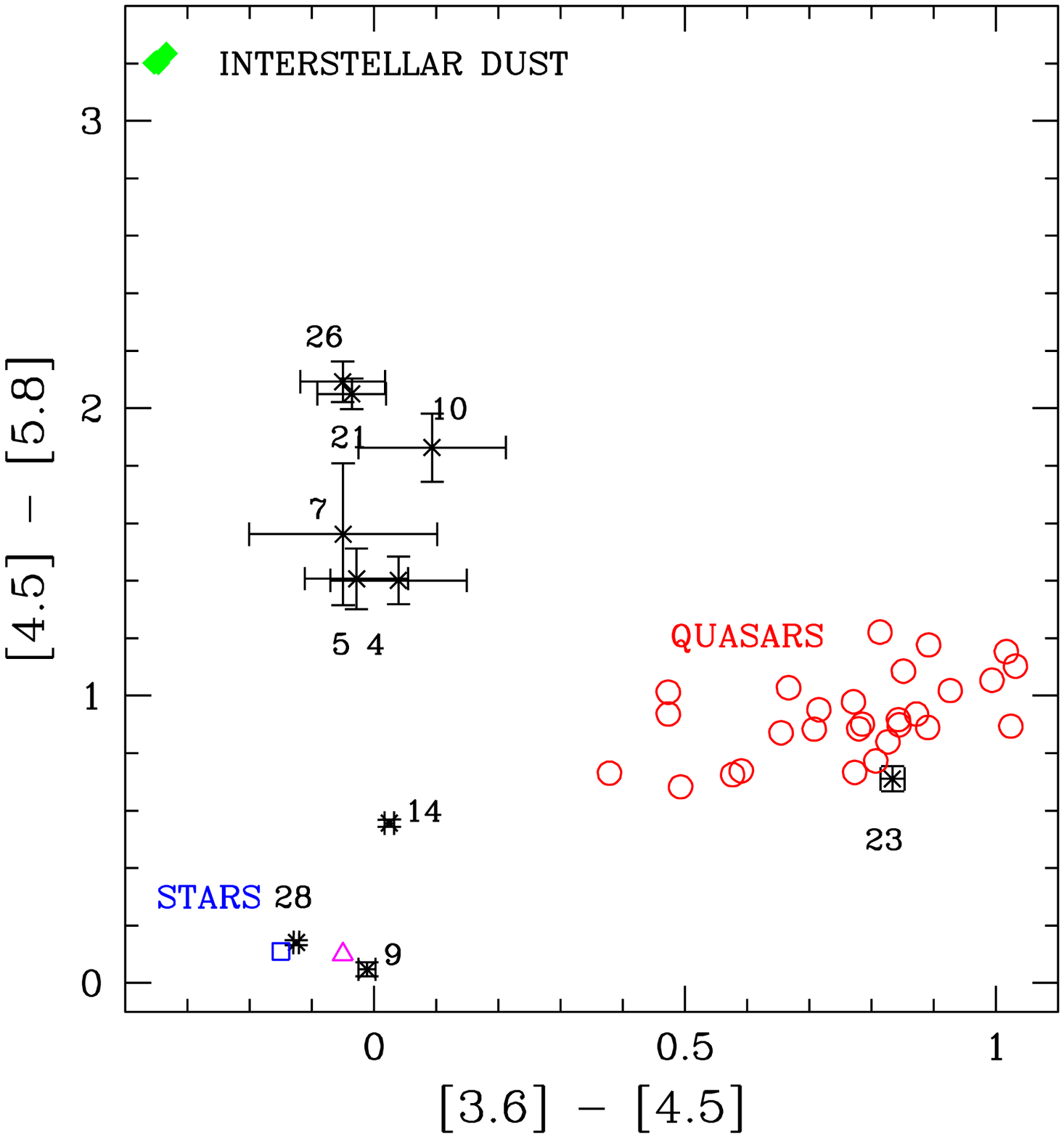}
\caption{
  \small 
The Spitzer IRAC 
[3.6] $-$ [4.5] vs. [4.5] $-$ [5.8] color-color
plot for the brightest 8 $\mu$m clumps in Arp 107 (black crosses).
The clumps are labeled.
The colors of
M0III stars (open dark blue square), from M. Cohen (2005, private 
communication), and the mean colors of the 
field stars of \citet{whitney04} (magenta triangle) are also shown.
The colors of normal stars all lie within 0.5 magnitudes of 0, 0 in this
plot (M. Cohen 2005, private communication).
The \citet{hatz05} colors of quasars are also plotted (red circles),
along with the colors
of the nuclei of the two
galaxies (\#14 and \#28).
The predicted IRAC colors for interstellar dust \citep{li01}
are also plotted (green diamonds), for ISRF 
strengths that vary from 0.3 $-$ 10,000 $\times$ that in the
solar neighborhood.  As the radiation field increases, the colors
become redder.  Note that the predicted dust colors vary very little
for this wide range in ISRF.
The errorbars include both statistical uncertainties and an uncertainty
in the colors due to varying the sky annuli (see text).
}
\end{figure}

 \begin{figure}
\plotone{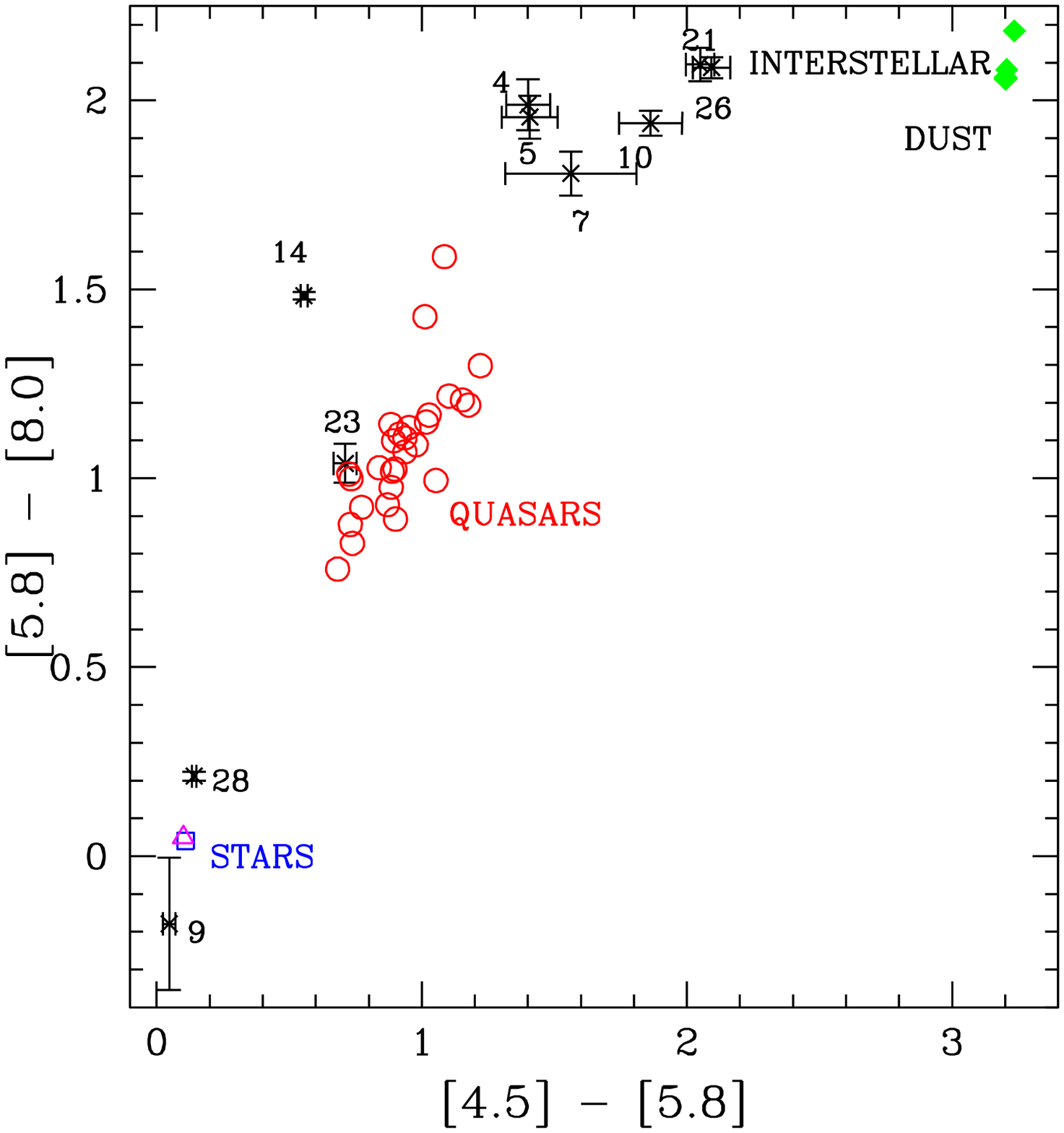}
\caption{
  \small 
The Spitzer IRAC [4.5] $-$ [5.8] vs. [5.8] $-$ [8.0] color-color
plot for the brightest 8 $\mu$m clumps in Arp 107 (black crosses).
The clumps are labeled.
The colors of
M0III stars (open dark blue square), from M. Cohen (2005, private 
communication), and the mean colors of the 
field stars of \citet{whitney04} (magenta triangle) are also shown.
The colors of normal stars all lie within 0.5 magnitudes of 0, 0 in this
plot (M. Cohen 2005, private communication).
The \citet{hatz05} colors of quasars are also plotted (red circles),
along with the colors
of the nuclei of the two
galaxies (\#14 and \#28).
The predicted IRAC colors for interstellar dust \citep{li01}
are also plotted (green diamonds), for ISRF 
strengths that vary from 0.3 $-$ 10,000 $\times$ that in the
solar neighborhood.  As the ISRF increases, the IRAC colors
of dust
become redder.  Note that the predicted dust colors vary very little
for this wide range in ISRF.
The errorbars include both statistical uncertainties and an uncertainty
in the colors due to varying the sky annuli (see text).
}
\end{figure}

 \begin{figure}
\plotone{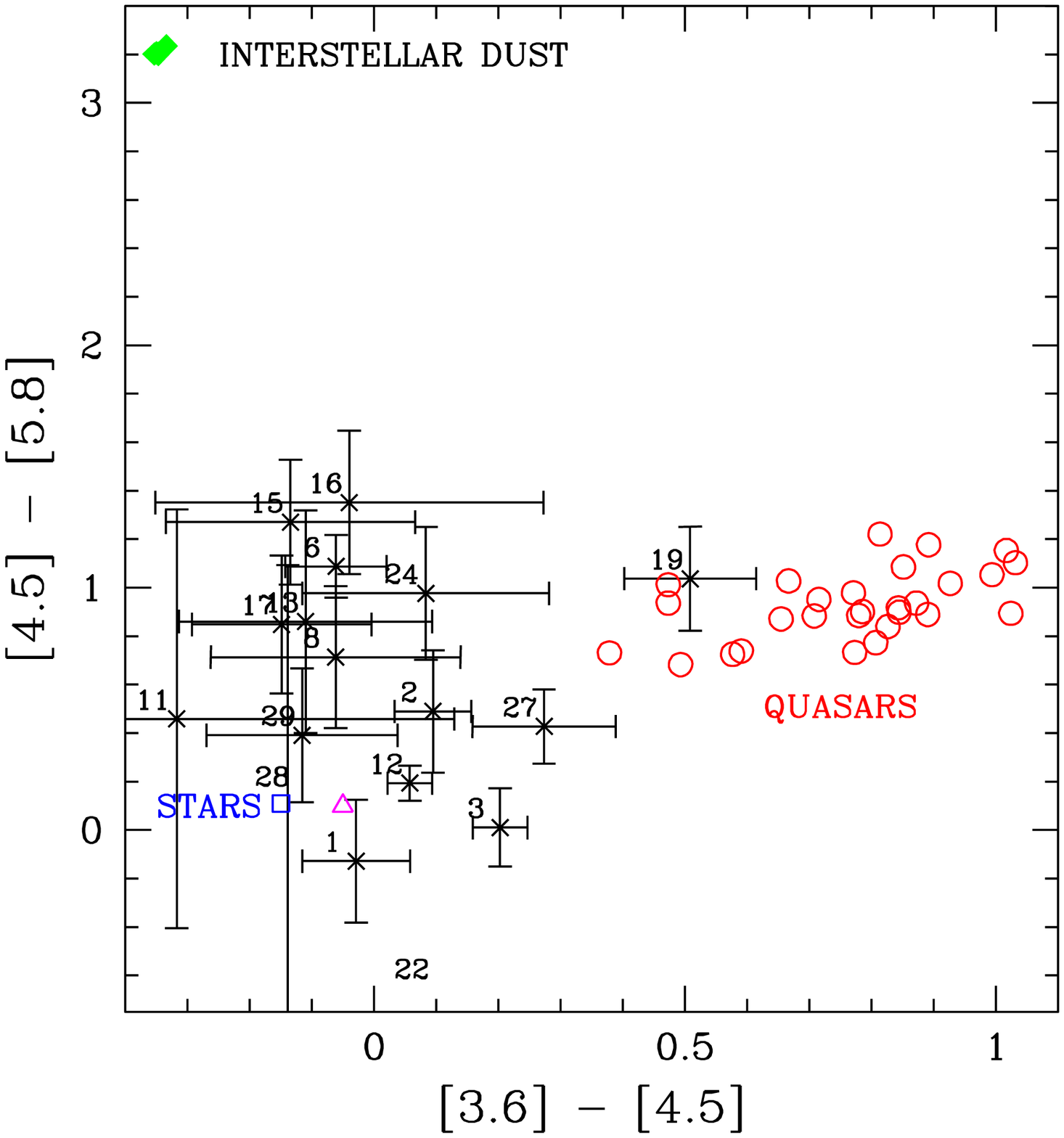}
\caption{
  \small 
The Spitzer 
IRAC [3.6] $-$ [4.5] vs. [4.5] $-$ [5.8] color-color
plot for
the fainter 
8 $\mu$m clumps in Figure 4 (black crosses).
The clumps are labeled.
The colors of
M0III stars (open dark blue square), from M. Cohen (2005, private 
communication), the mean colors of the field stars of \citet{whitney04} (magenta triangle),
the predicted IRAC colors for interstellar dust \citep{li01} (green diamonds),
and the \citet{hatz05} quasars colors (red circles)
are also plotted.
The very faint sources \#18 and \#25 are not plotted since their
uncertainties are very large and their colors are not well-constrained.
}
\end{figure}

 \begin{figure}
\plotone{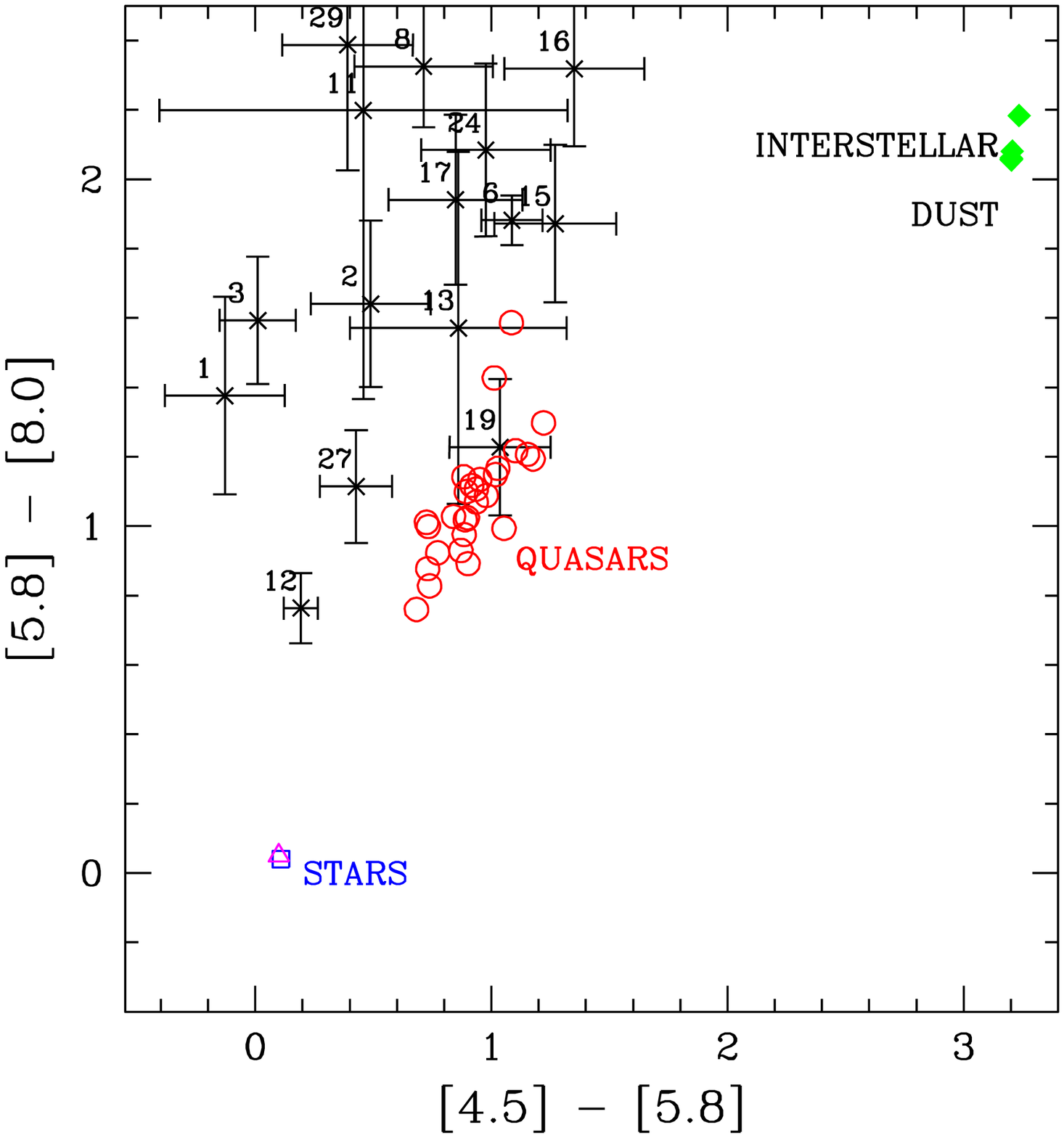}
\caption{
  \small 
The Spitzer 
IRAC [4.5] $-$ [5.8] vs. [5.8] $-$ [8.0] color-color
plot for
the fainter 
8 $\mu$m clumps in Figure 4 (black crosses).
Some of the clumps are labeled.
The colors of
M0III stars (open dark blue square), from M. Cohen (2005, private 
communication), the mean colors of the field stars of \citet{whitney04} (magenta triangle),
the predicted IRAC colors for interstellar dust \citep{li01} (green diamonds),
and the \citet{hatz05} quasars colors (red circles)
are also plotted.
The very faint sources \#18, \#22, and \#25 are not plotted since their
uncertainties are very large and their colors are not well-constrained.
}
\end{figure}

 \begin{figure}
\plotone{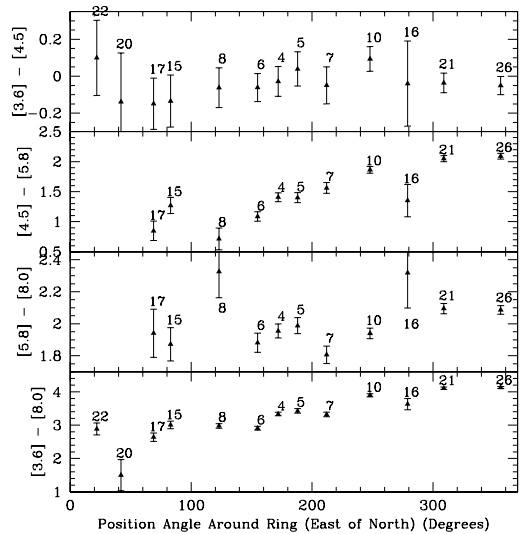}
\caption{
  \small 
The Spitzer IRAC colors for the clumps in the ring/arm, as a function of position angle.
Clumps \#20 and \#22 are not included in the second and third panels
since the 5.8 $\mu$m uncertainties are very large.
}
\end{figure}

 \begin{figure}
\plotone{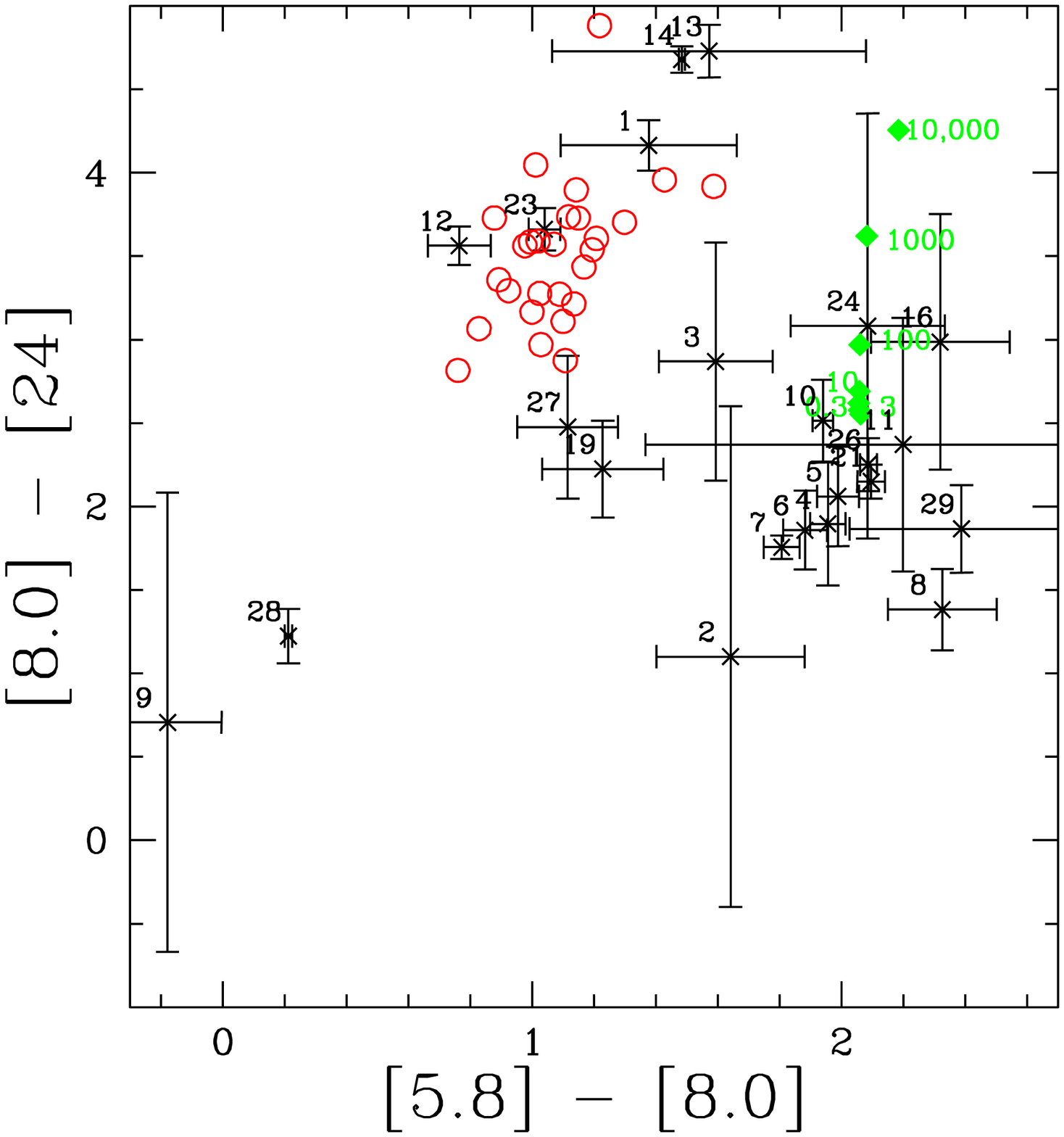}
\caption{
  \small 
For the identified clumps, 
the Spitzer IRAC [5.8] $-$ [8.0] color plotted
against the  IRAC-MIPS [8.0] $-$ [24] color (black crosses).
The clumps are labeled to the upper left of the corresponding
data point.
The colors of
the \citet{hatz05} quasars colors (red circles)
are also plotted, along with the predicted colors
for 
interstellar dust \citep{li01} (green diamonds).
The green labels on the data points for dust correspond
to the ISRF in terms of that in
the solar neighborhood.
The errorbars include both statistical uncertainties as well
as an extra term due to background region selection, as discussed in
the text.
For clarity, clumps \#15, 17, 20, and 22 are omitted because they
have very large errorbars.
}
\end{figure}

 \begin{figure}
\plotone{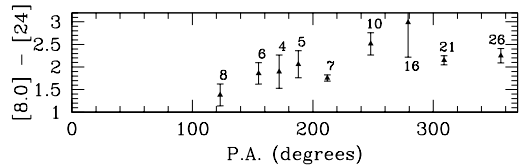}
\caption{
  \small 
For the clumps in the ring/arm, the 
Spitzer [8.0] $-$ [24] color as a function of position angle.
For clarity, clumps \#15, 17, 20, and 22 are not plotted, since their
errorbars are very large.
}
\end{figure}

 \begin{figure}
\plotone{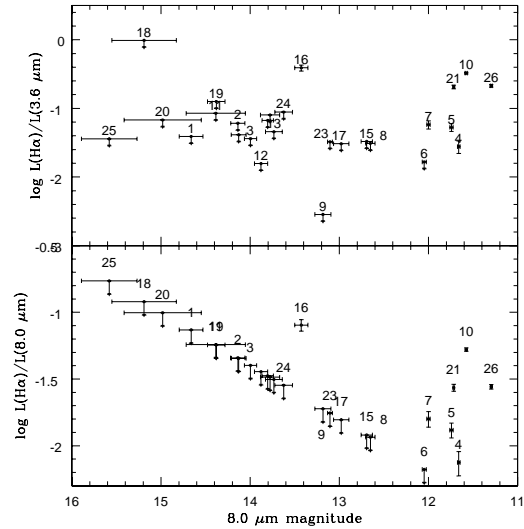}
\caption{
  \small 
Top:
For the identified clumps, 
the ratio of H$\alpha$ luminosity to 3.6 $\mu$m luminosity, as a function of 8.0 $\mu$m
magnitude.
Some of the clumps are labeled.
Bottom:
the ratio of H$\alpha$ luminosity to 8.0 $\mu$m luminosity, as a function of 8.0 $\mu$m
magnitude.
The upper limits are 3$\sigma$.
}
\end{figure}

 \begin{figure}
\plotone{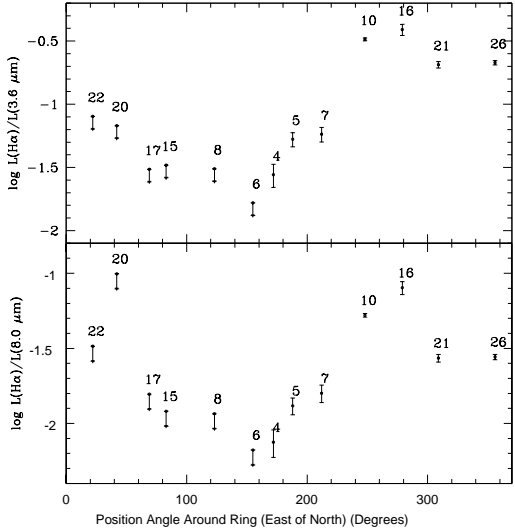}
\caption{
  \small 
Top: the ratio of H$\alpha$ luminosity to 3.6 $\mu$m luminosity
as a function of position angle, for the clumps in the ring/arm.
Bottom: the ratio of H$\alpha$ luminosity to 8.0 $\mu$m luminosity
as a function of position angle, for the clumps in the ring/arm.
The plotted errorbars only include statistical errors,
not uncertainties in the absolute photometry.
}
\end{figure}

 \begin{figure}
\caption{
  \small 
Four snapshots from a numerical hydrodynamical model of the Arp 107 system. 
The top left panel shows a sky plane view of gas and star particles at a 
time near the present. The dashed line shows the relative orbit of the 
companion from the time of the first close approach. The top right panel 
is the same, but viewed from an orthogonal perspective. The lower panels 
show curves of test particles that were circular rings in the initial 
primary disk, at two earlier times, in the sky plane. In both cases the 
extension to the west is made up of particles in the bridge or captured 
onto the companion. These two panels show the development of 
compression at the front of the primary arm, which is the likely 
driver of the observed star formation in this system. 
Length units are in kpc, with the adopted scaling.  See text for details.
}
\end{figure}

\clearpage


\clearpage


\clearpage


\clearpage








%
%
\begin{deluxetable}{rrrrrrrrrrrrrrr}
\tabletypesize{\scriptsize}
\def\et#1#2#3{${#1}^{+#2}_{-#3}$}
\tablewidth{0pt}
\tablecaption{Spitzer Clumps in the Arp 107 Field\label{tab-1}}
\tablehead{
\colhead{ID} &
\multicolumn{3}{c}{R.A.}&
\multicolumn{3}{c}{Dec.}&
\colhead{F$_{3.6 {\mu}m}$} & 
\colhead{F$_{4.5 {\mu}m}$} & 
\colhead{F$_{5.8 {\mu}m}$} & 
\colhead{F$_{8.0 {\mu}m}$} & 
\colhead{F$_{24 {\mu}m}$} & 
\colhead{F$_{H{\alpha}+[N~II]}$}
\\ 
&
\multicolumn{6}{c}{(J2000)}
& \multicolumn{1}{c}{(mJy)} 
& \multicolumn{1}{c}{(mJy)} 
& \multicolumn{1}{c}{(mJy)} 
& \multicolumn{1}{c}{(mJy)} 
& \multicolumn{1}{c}{(mJy)} 
& \multicolumn{1}{c}{(10$^{-15}$} 
\\ 
&
\multicolumn{6}{c}{}
& \multicolumn{1}{c}{} 
& \multicolumn{1}{c}{} 
& \multicolumn{1}{c}{} 
& \multicolumn{1}{c}{} 
& \multicolumn{1}{c}{} 
& \multicolumn{1}{c}{erg~s$^{-1}$ cm$^{-2}$)} 
}
\startdata
 1  &  10  &  52  &  15.8  &   30  &   2  &  42.9  &       0.123  &       0.078  &       0.045  &       0.086  &        0.461  &  $\le$1.0   \\
 2  &  10  &  52  &  14.0  &   30  &   2  &  43.5  &       0.079  &       0.056  &       0.057  &       0.140  &  $\le$0.044  &  $\le$1.0   \\
 3  &  10  &  52  &  17.8  &   30  &   2  &  51.3  &       0.132  &       0.103  &       0.068  &       0.159  &        0.258  &  $\le$1.0   \\
 4  &  10  &  52  &  15.4  &   30  &   2  &  56.7  &       0.279  &       0.176  &       0.417  &       1.367  &        0.907  &        1.7   \\
 5  &  10  &  52  &  14.7  &   30  &   2  &  58.1  &       0.237  &       0.159  &       0.375  &       1.268  &        0.979  &        2.7   \\
 6  &  10  &  52  &  15.9  &   30  &   3  &   0.2  &       0.289  &       0.177  &       0.313  &       0.957  &        0.613  &  $\le$1.0   \\
 7  &  10  &  52  &  13.6  &   30  &   3  &   2.6  &       0.207  &       0.128  &       0.350  &       1.001  &        0.584  &        2.6   \\
 8  &  10  &  52  &  17.0  &   30  &   3  &  11.7  &       0.155  &       0.095  &       0.119  &       0.548  &        0.227  &  $\le$1.0   \\
 9  &  10  &  52  &  14.1  &   30  &   3  &  12.7  &       1.685  &       1.079  &       0.732  &       0.336  &  $\le$0.075  &  $\le$1.0   \\
10  &  10  &  52  &  12.8  &   30  &   3  &  17.3  &       0.180  &       0.127  &       0.458  &       1.478  &        1.733  &       12.8   \\
11  &  10  &  52  &  16.3  &   30  &   3  &  18.3  &       0.057  &       0.027  &       0.027  &       0.111  &        0.114  &  $\le$1.0   \\
12  &  10  &  52  &  18.6  &   30  &   3  &  21.2  &       0.306  &       0.209  &       0.162  &       0.177  &        0.546  &  $\le$1.0   \\
13  &  10  &  52  &  14.0  &   30  &   3  &  26.7  &       0.105  &       0.061  &       0.088  &       0.202  &        1.824  &  $\le$1.0   \\
14  &  10  &  52  &  15.0  &   30  &   3  &  28.4  &       2.793  &       1.848  &       2.003  &       4.250  &       36.603  &  $\le$1.0   \\
15  &  10  &  52  &  17.2  &   30  &   3  &  29.0  &       0.146  &       0.083  &       0.174  &       0.528  &        0.584  &  $\le$1.0   \\
16  &  10  &  52  &  12.6  &   30  &   3  &  32.3  &       0.042  &       0.026  &       0.059  &       0.269  &        0.486  &        3.6   \\
17  &  10  &  52  &  16.9  &   30  &   3  &  37.7  &       0.157  &       0.088  &       0.125  &       0.406  &        0.679  &  $\le$1.0   \\
18  &  10  &  52  &  13.6  &   30  &   3  &  39.0  &       0.005  &  $\le$0.001  &  $\le$0.005  &       0.053  &        0.258  &  $\le$1.0   \\
19  &  10  &  52  &  10.9  &   30  &   3  &  45.1  &       0.038  &       0.039  &       0.067  &       0.112  &        0.100  &  $\le$1.0   \\
20  &  10  &  52  &  16.2  &   30  &   3  &  47.1  &       0.071  &       0.040  &  $\le$0.006  &       0.064  &        0.290  &  $\le$1.0   \\
21  &  10  &  52  &  12.8  &   30  &   3  &  49.1  &       0.130  &       0.081  &       0.348  &       1.298  &        1.087  &        5.8   \\
22  &  10  &  52  &  15.7  &   30  &   3  &  54.3  &       0.060  &       0.042  &       0.015  &       0.195  &        0.446  &  $\le$1.0   \\
23  &  10  &  52  &   9.9  &   30  &   3  &  56.8  &       0.147  &       0.205  &       0.256  &       0.361  &        1.219  &  $\le$1.0   \\
24  &  10  &  52  &  15.8  &   30  &   4  &   1.1  &       0.054  &       0.038  &       0.061  &       0.224  &        0.443  &  $\le$1.0   \\
25  &  10  &  52  &  11.3  &   30  &   4  &   1.5  &       0.133  &       0.092  &  $\le$0.005  &       0.037  &  $\le$0.041  &  $\le$1.0   \\
26  &  10  &  52  &  14.7  &   30  &   4  &   5.3  &       0.188  &       0.116  &       0.517  &       1.913  &        1.760  &        8.7   \\
27  &  10  &  52  &  17.6  &   30  &   4  &   8.1  &       0.116  &       0.097  &       0.093  &       0.141  &        0.159  &  $\le$1.0   \\
28  &  10  &  52  &  18.5  &   30  &   4  &  20.9  &       5.457  &       3.145  &       2.326  &       1.530  &        0.546  &  $\le$1.0   \\
29  &  10  &  52  &  12.6  &   30  &   4  &  21.3  &       0.072  &       0.042  &       0.039  &       0.190  &        0.122  &  $\le$1.0   \\
\enddata
\tablenotetext{a}{The statistical uncertainties at 3.6, 4.5, 5.8, 8.0,
and 24 $\mu$m are $\le$0.002, $\le$0.001, $\le$0.002, $\le$0.005, and 
$\le$0.013 mJy, respectively, determined from the rms in the sky
annuli.  The absolute calibration uncertainty is $\le$10$\%$ (IRAC
Data Handbook).
}
\tablenotetext{b}{The statistical 
H$\alpha$+[N~II] uncertainties
are $\approx$3.5 $\times$ 10$^{-16}$ erg~s$^{-1}$ cm$^{-2}$.  The absolute
H$\alpha$+[N~II] calibration uncertainty is $\approx$30$\%$.
}
\end{deluxetable}
%



\end{document}